\documentclass{SciPost}

\hypersetup{
    colorlinks,
    linkcolor={red!50!black},
    citecolor={blue!50!black},
    urlcolor={blue!80!black}
}

\usepackage[bitstream-charter]{mathdesign}
\DeclareSymbolFont{usualmathcal}{OMS}{cmsy}{m}{n}
\DeclareSymbolFontAlphabet{\mathcal}{usualmathcal}

\fancypagestyle{SPstyle}{
\fancyhf{}
\lhead{\colorbox{scipostblue}{\bf \color{white} ~SciPost Physics Proceedings }}
\rhead{{\bf \color{scipostdeepblue} ~Submission }}

\fancyfoot[C]{\textbf{\thepage}}
}
 
\newcommand{\ttbar}{\ensuremath{t\bar{t}}}

\newcommand{\pT}{\ensuremath{p_{\mathrm{T}}}}

\begin{document}

\pagestyle{SPstyle}

\begin{center}{\Large \textbf{\color{scipostdeepblue}{
Top quark spin and quantum entanglement\\ in the ATLAS experiment\\
}}}\end{center}

\begin{center}\textbf{
Roman Lys\'ak\textsuperscript{$\star$} on behalf of the ATLAS Collaboration~\footnote{Copyright CERN for the benefit of the ATLAS Collaboration. CC-BY-4.0 license}
}\end{center}

\begin{center}
FZU - Institute of Physics of the Czech Academy of Sciences, Prague, Czech Republic
\\[\baselineskip]
$\star$ \href{mailto:roman.lysak@cern.ch}{\small roman.lysak@cern.ch}
\end{center}

\definecolor{palegray}{gray}{0.95}
\begin{center}
\colorbox{palegray}{
  \begin{tabular}{rr}
  \begin{minipage}{0.36\textwidth}
    \includegraphics[width=60mm,height=1.5cm]{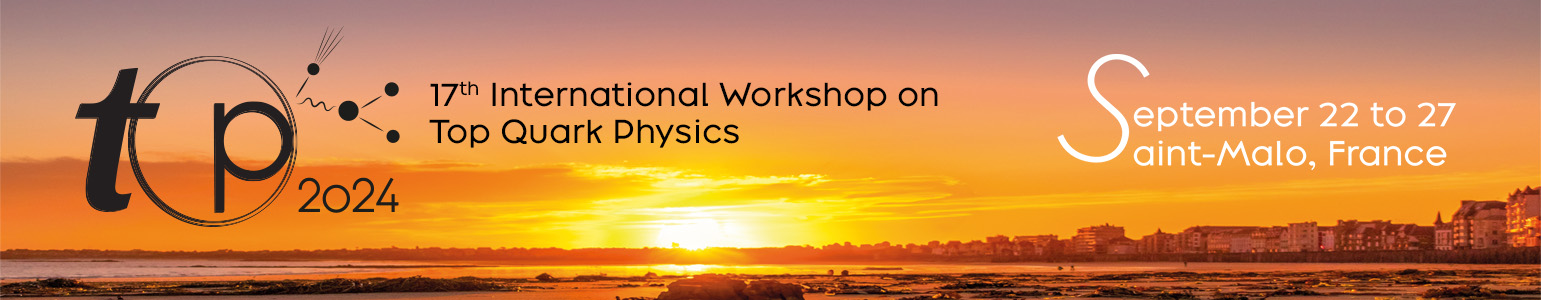}
  \end{minipage}
  &
  \begin{minipage}{0.55\textwidth}
    \begin{center} \hspace{5pt}
    {\it The 17th International Workshop on\\ Top Quark Physics (TOP2024)} \\
    {\it Saint-Malo, France, 22-27 September 2024
    }
    \doi{10.21468/SciPostPhysProc.?}\\
    \end{center}
  \end{minipage}
\end{tabular}
}
\end{center}

\section*{\color{scipostdeepblue}{Abstract}}
\textbf{\boldmath{%
    Top quark pair spin correlation measurements performed by the ATLAS experiment using $pp$ collisions at the CERN Large Hadron Collider are summarized. Moreover, the measurement of a specific observable $D$ related to top quark pair spin correlations is presented using the full LHC Run 2 data taking at the center-of-mass energy of $\sqrt{s} = 13$~TeV. This allowed the ATLAS experiment to observe the quantum entanglement, one of the fundamental property of the quantum mechanics.
}}

\vspace{\baselineskip}

\noindent\textcolor{white!90!black}{%
\fbox{\parbox{0.975\linewidth}{%
\textcolor{white!40!black}{\begin{tabular}{lr}%
  \begin{minipage}{0.6\textwidth}%
    {\small Copyright attribution to authors. \newline
    This work is a submission to SciPost Phys. Proc. \newline
    License information to appear upon publication. \newline
    Publication information to appear upon publication.}
  \end{minipage} & \begin{minipage}{0.4\textwidth}
    {\small Received Date \newline Accepted Date \newline Published Date}%
  \end{minipage}
\end{tabular}}
}}
}




\section{Introduction}
\label{sec:intro}
The top quark differs from other quarks mainly by its large mass $m_t = 172.57 \pm 0.29$~GeV~\cite{ParticleDataGroup:2024cfk}. Consequently, the top quark decays well before forming bound states and before spin decorrelation effects happen~\cite{Bigi:1986jk}. Therefore, the correlation between the spin of the top quark and the antitop quark in their pair (\ttbar) production is transferred to their decay products. By measuring these products, it is thus possible to learn about these correlations. Recently, it has been proposed that measuring  \ttbar\ correlations in a specific phase space can be used to study the fundamental aspects of quantum mechanics at the LHC~\cite{Afik:2020onf,Afik:2022kwm}. 

The ATLAS experiment~\cite{ATLAS:2008xda} performed a few measurements of \ttbar\ spin correlations using the data from LHC Run 1 at the center-of-mass energy $\sqrt{s} = 7$~TeV~\cite{TOPQ-2011-11,TOPQ-2013-01,TOPQ-2013-10} and $\sqrt{s} = 8$~TeV~\cite{TOPQ-2014-07,TOPQ-2015-13} and also from LHC Run 2 at $\sqrt{s} = 13$~TeV~\cite{TOPQ-2016-10}. 

In this paper, we summarize the most precise spin density matrix ATLAS measurement performed at $\sqrt{s} = 8$~TeV~\cite{TOPQ-2015-13} and also a simplified measurement of \ttbar\ spin correlations at $\sqrt{s} = 13$~TeV~\cite{TOPQ-2016-10}. Moreover, the measurement of a specific observable $D$ sensitive to \ttbar\ spin correlations and used to determine the quantum entanglement is also presented~\cite{ATLAS:2023fsd}.

\section{Top quark pair spin correlations}

The spin information of top quarks in the \ttbar\ pair production can be fully characterized by 15 coefficients where six coefficients correspond to three-component polarization vectors of top ($B_{+}$) and antitop ($B_{-}$) quarks while nine coefficients ($3 \times 3$ matrix $C$) describe the correlations between the spin of the top and the antitop quarks alongside three axes~\cite{Bernreuther:2015yna}.
The orthonormal helicity basis with three axes $\vec{k},\vec{n},\vec{r}$ is typically used to define these coefficients. All these coefficients can be determined by measuring various $\cos{\theta}^{a}_{+/-}$ angles between the direction of a positive(+) or negative(-) lepton in its parent top quark rest frame and a given axis $a$: $B^{a}_{+/-}= 3 <\cos \theta^a_{+/-}>$, $C(a,b)= -9<\cos \theta^{a}_{+}\cos\theta^{b}_{-}>$, where $a$,$b$ correspond to one of $\vec{k},\vec{n},\vec{r}$ axis~\cite{Bernreuther:2015yna}.

\begin{figure*}[htbp]
  \centering
  \includegraphics[width=0.495\textwidth]{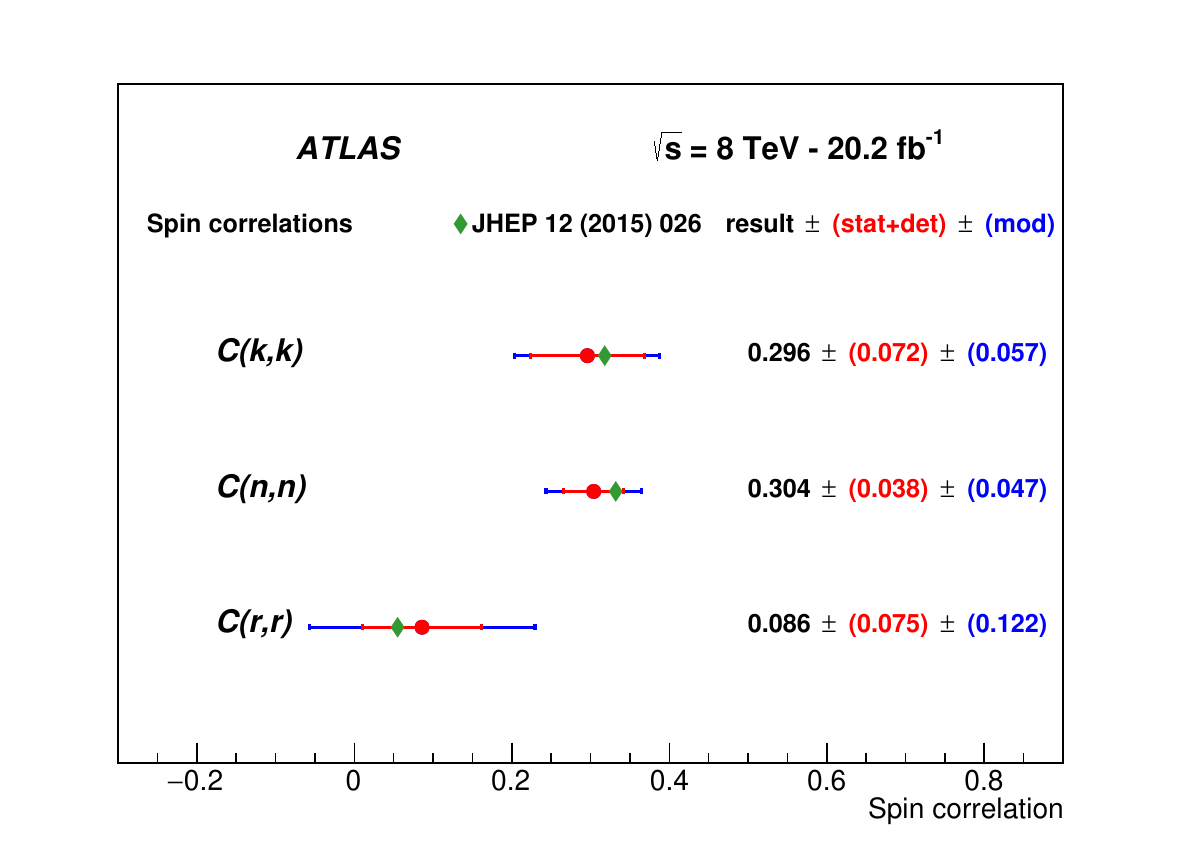}
  \includegraphics[width=0.495\textwidth]{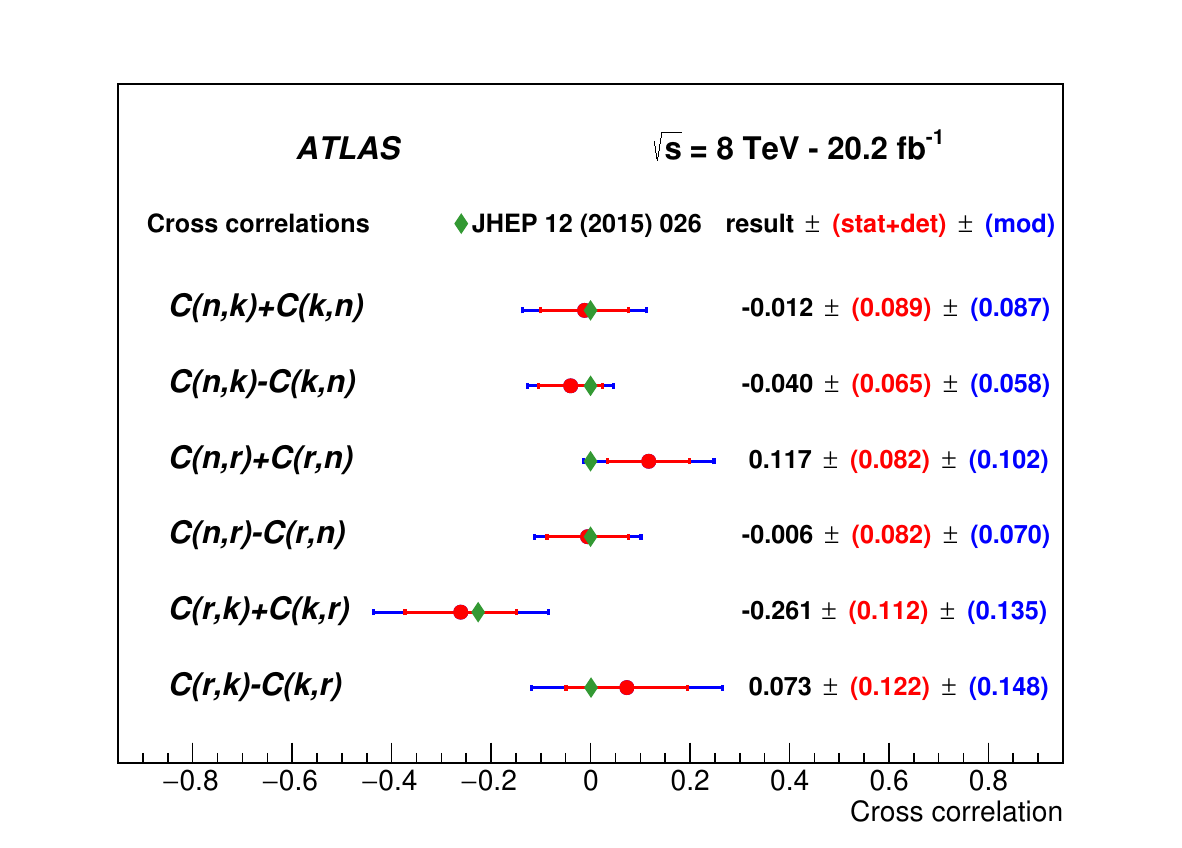}
  \caption{
    Comparison of the measured spin correlations (data points) with predictions from the Standard Model (diamonds) for the parton-level measurement. The spin correlations along the same axes are present in the left plot, while the cross-correlation terms using different axes are presented on the right. Inner bars indicate statistical and detector related uncertainties, outer bars indicate modeling systematics, summed in quadrature~\cite{TOPQ-2015-13}.
  }
  \label{fig:spinCorr8TeV}
\end{figure*}

The ATLAS experiment measurement of the full spin density matrix was performed in \mbox{Run 1} at $\sqrt{s} = 8$~TeV using the full dataset ($20.2\ {\rm fb}^{-1}$)~\cite{TOPQ-2015-13}. 
All three dilepton channels ($ee,\mu\mu,e\mu$) were combined. The neutrino weighting method was used for $t\bar{t}$ reconstruction. Various measured $\cos\theta$ distributions were corrected to the truth level (both the stable-particle level and the parton level) and the coefficients were determined from such distributions. The measured polarization coefficients are all consistent with the Standard Model (SM) prediction of zero. The summary of measured correlation coefficients is presented in Figure~\ref{fig:spinCorr8TeV}. Most of the measurements agree within one standard deviation with the theory. The uncertainties range from $3-5\%$ for spin polarization measurements up to $9-19\%$ for spin correlation cross-terms.

\begin{figure*}[tbp]
  \centering
  \includegraphics[width=0.48\textwidth]{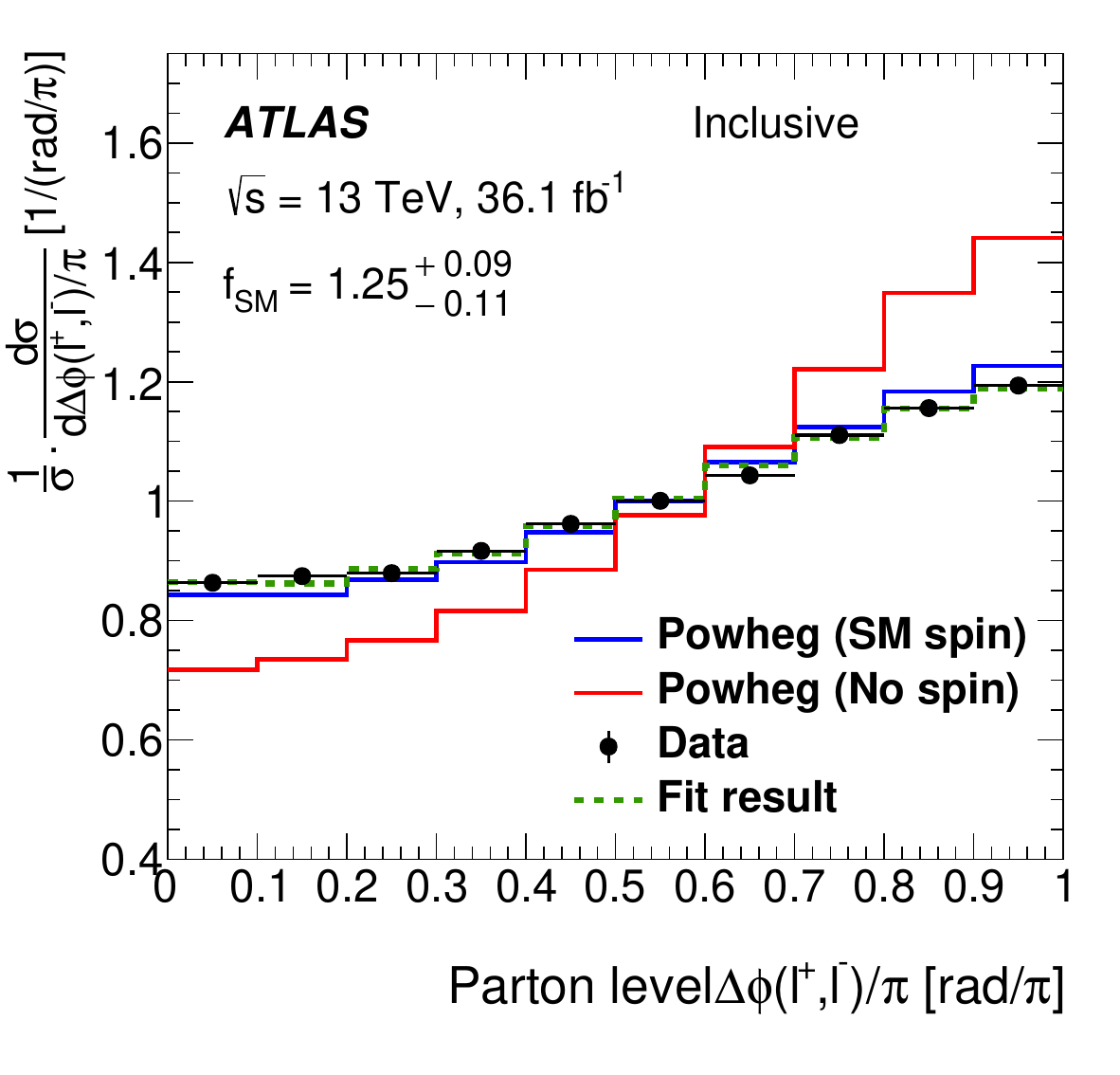} 
  \caption{
    Result of the fit of hypothesis templates corresponding to SM spin correlation and no-correlation to the unfolded data showing the $\Delta\phi(l^+,l^-)$ distribution~\cite{TOPQ-2016-10}.
  }
  \label{fig:spinCorr13TeV}
\end{figure*}

The ATLAS experiment performed also measurement of spin correlations with partial Run 2 dataset ($36\ {\rm fb}^{-1}$) at $\sqrt{s} = 13$~TeV~\cite{TOPQ-2016-10}. Here, the $e\mu$ channel was used and only the difference in the azimuthal angle between the leptons $\Delta\phi(l^+,l^-)$ was measured. The template fit of data to the prediction, shown in Figure~\ref{fig:spinCorr13TeV},  was performed to determine the amount of spin correlations compared to the SM prediction obtained at next-to-leading order using the {\sc POWHEG} Monte-Carlo (MC) generator. The measured fraction was determined to be $1.25 ^{+0.09}_{-0.11}$ which corresponds to a bit higher correlations than the prediction from {\sc POWHEG}. This difference gets smaller if the next-to-next-to-leading order prediction is used in comparison~\cite{Behring:2019iiv}.

\section{Quantum entanglement} 
The quantum entanglement is a phenomenon in which a quantum state of one particle can not be described independently from another particle. A typical example of the entangled state is a system of two fermions in a spin-singlet state. At the LHC, the initial state is not a pure quantum state, it's a mixed spin state of initial quarks or gluons. In general, such system is described by a density matrix. This corresponds to a \ttbar\ spin density matrix $\rho$ when interested in the spins of top quarks in their pair production. The spin correlations can be then used to characterize the entanglement. 

A quantitative measure of the degree of the entanglement is a concurrence of the spin density matrix $C[\rho]$. The sufficient and necessary condition for the entanglement is $C[\rho] > 0$~\cite{Afik:2020onf}. There are two distinct regions with high values of a concurrence in the two dimensional plane of the invariant mass of the \ttbar\ pair $m(\ttbar)$ and the production angle $\theta$ in the \ttbar\ center-of-mass frame: a region of high $m(\ttbar)$ while $\theta \sim \pi/2$ and a low $m(\ttbar)$ region. The dominant contribution in a low mass region is the $gg$ fusion with top quarks in the spin singlet state. The sufficient condition on the concurrence can be translated to the sufficient condition on \ttbar\ spin correlation matrix in a low mass region $Tr[C] < -1$~\cite{Afik:2020onf}. The entanglement can be observed using $D$ observable which is directly proportional to a slope of the $\cos \phi$ distribution $D = Tr[C]/3 = -3⟨ \cos \phi ⟩$, where $\phi$ is an angle between the two lepton directions measured in their parent top quark and antiquark rest frames, respectively. The entanglement condition $Tr[C] < -1$ then translates to $D < -1/3$~\cite{Afik:2020onf}. 

The ATLAS experiment performed the measurement in the dilepton $e\mu$ channel using the full \mbox{Run 2} dataset corresponding to the luminosity of $140\ {\rm fb}^{-1}$ collected between years 2015 and 2018~\cite{ATLAS:2023fsd}. In this measurement, the ATLAS collaboration concentrated on a low $m(\ttbar)$ region. The simple selection criteria were used. One electron and one muon with $\pT > 25-28$~GeV (depending on the year of data taking) were required with the opposite-sign electric charge. No cut on missing transverse momentum was applied while at least two jets with $\pT > 25$~GeV were required where at least one of them was identified to come from the hadronization of a $b$-hadron ($b$-jets). There are about 1.1 million of events after full event selection with ~90\% of these expected to come from the \ttbar\ pair signal process. The dominant background is the single top production (60\%) and events with non-prompt or mis-identified lepton (`fakes', 30\%). All backgrounds except for fakes are estimated by the simulation with fakes estimated by the data-driven method.

The reconstruction of top quarks momenta was performed by a combination of various methods. The ‘Ellipse’ method (85\% efficiency) was used as the main method. 
This method analytically calculates two ellipses for $\pT$ of neutrinos and finds intersections. If ‘Ellipse’ method fails, the neutrino weighting method (5\%) is used while if both methods fail, a simple pairing of leptons with the closest b-jets is performed (10\%).

\begin{figure*}[tbp]
  \centering
  \includegraphics[width=0.48\textwidth]{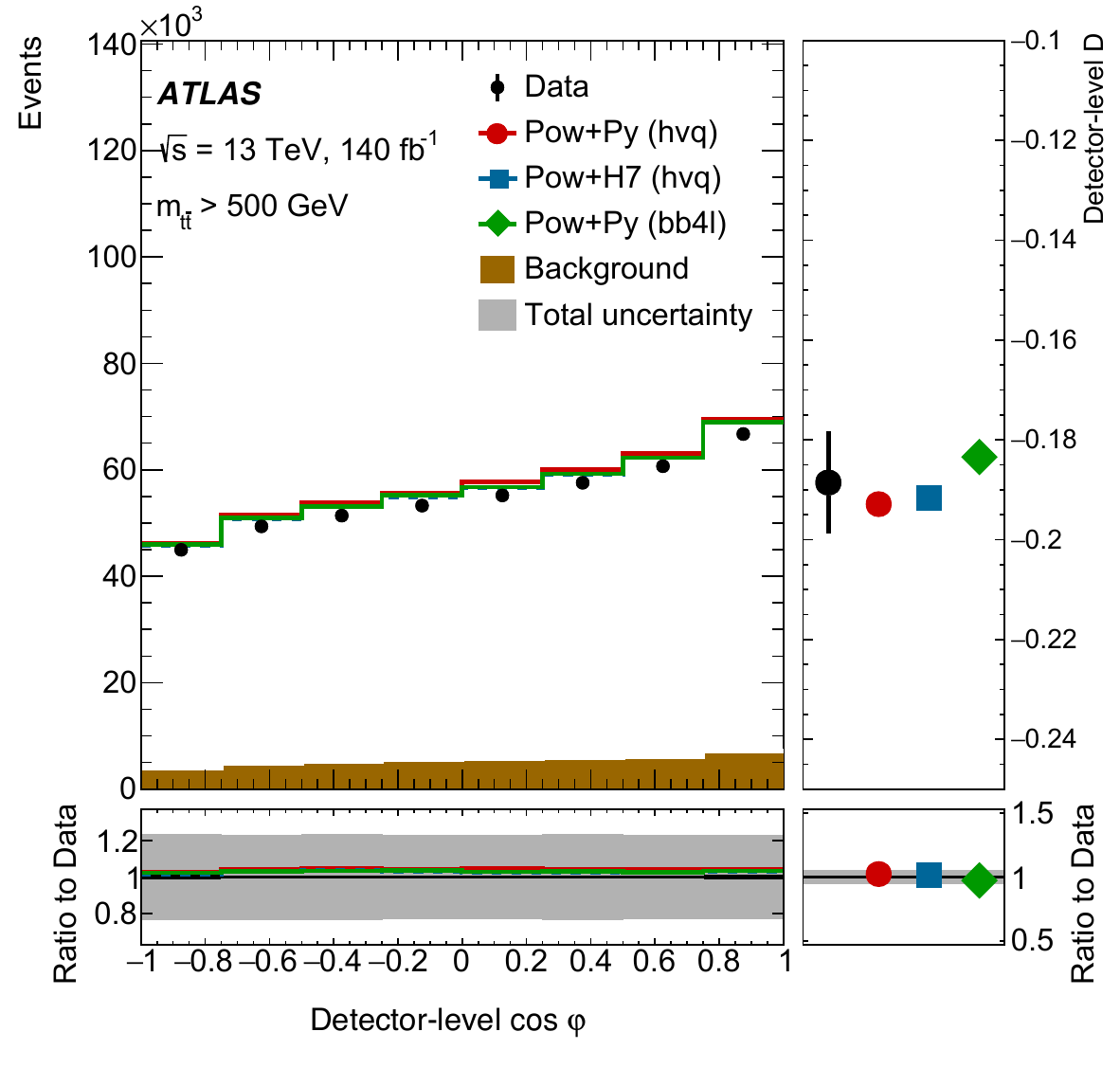}
  \includegraphics[width=0.48\textwidth]{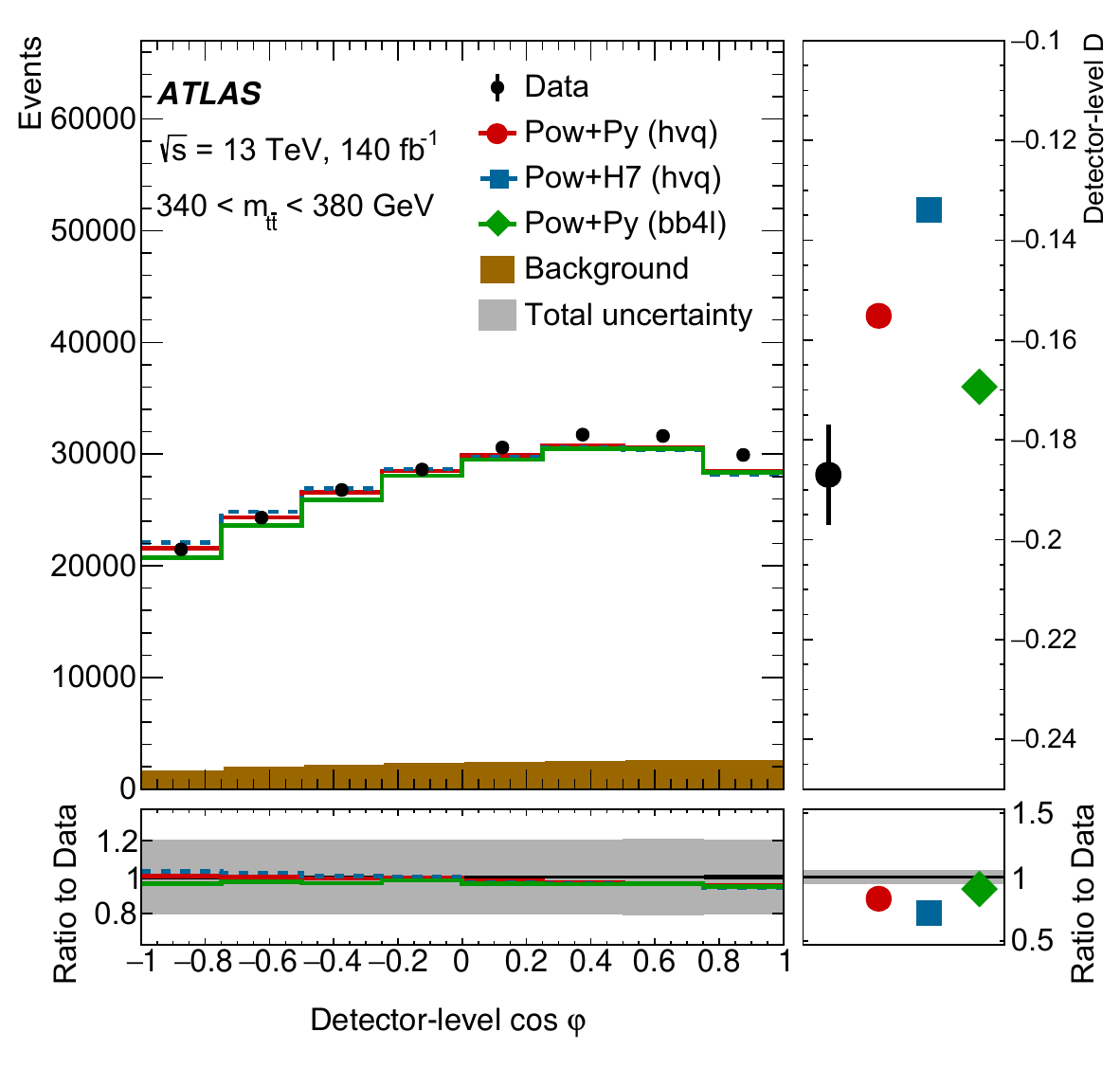}
  \caption{
    The left panel shows the $\cos\phi$ distribution at the detector level and the right panel shows the entanglement marker $D$. The left plot corresponds to the validation region and the right plot to the signal region at the detector level. Three different MC generator predictions are shown after background processes are subtracted. The uncertainty band shows the uncertainties from all sources added in quadrature~\cite{ATLAS:2023fsd}. }
  \label{fig:entanglement:reco_level}
\end{figure*}

The event sample was divided into 3 regions based on the $m(\ttbar)$: in the signal region it is required $340\ {\rm GeV} < m(\ttbar) < 380\ {\rm GeV}$ while two validation regions require $m(\ttbar) > 500$~GeV and $380\ {\rm GeV} < m(\ttbar) < 500\ {\rm GeV}$, respectively. The reconstructed detector level $\cos \phi$ distribution and the corresponding observable $D$ agree well with the prediction in both validation regions, see Figure~\ref{fig:entanglement:reco_level} for $m(\ttbar) > 500$~GeV region. The detector level $\cos \phi$ distribution and the observable $D$ in the signal region is also shown in Figure~\ref{fig:entanglement:reco_level}. 

The measured data at the detector level are corrected to the truth (stable-particle) level using the calibration curve where the particle level selection criteria are similar to the detector level. The calibration curve was created by reweighting the simulation based on the truth $D$ value. For each systematic uncertainty,  a new calibration curve is created. There are three main categories of systematic uncertainties: the signal modeling, which is the dominant component (3.2\%); the background modeling (1.1\%); and the object reconstruction. The main component of the signal modeling is the uncertainty due to the modeling of the top quark decay (1.6\%).

\begin{figure*}[tbp]
  \centering
  \includegraphics[width=0.6\textwidth]{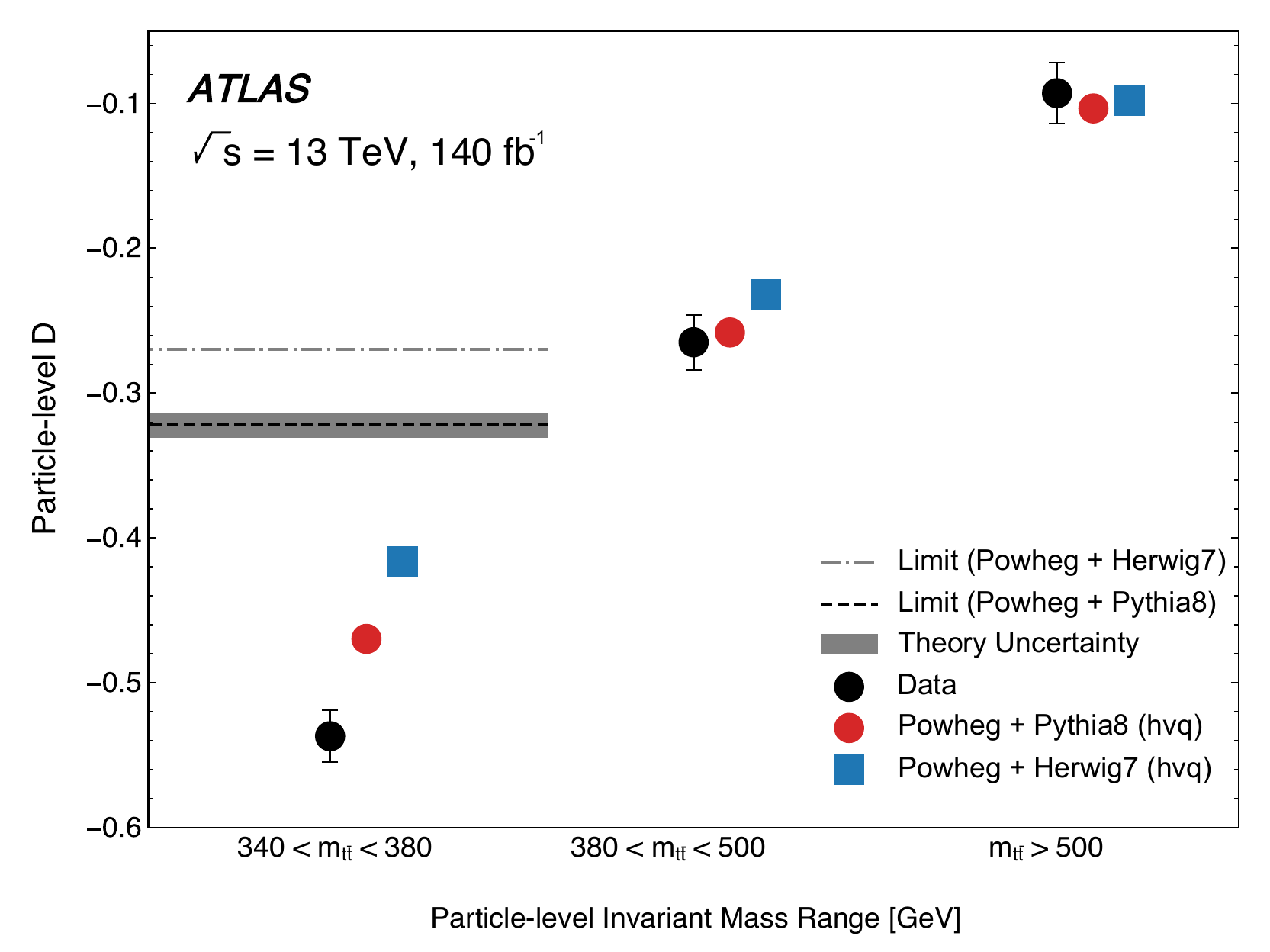}
  \caption{
    The particle-level $D$ results in the signal and validation regions compared with various predictions from Monte-Carlo generators. The entanglement limit shown is a conversion from its parton-level value of $D = -1/3$ to the corresponding value at particle level~\cite{ATLAS:2023fsd}.
  }
  \label{fig:entanglement:signal_region}
\end{figure*}

Measured $D$ in the signal region at the particle level is $-0.537 \pm 0.002\ {\rm (stat.)} \pm 0.019\ {\rm (syst.)}$, see Figure~\ref{fig:entanglement:signal_region}, which is significantly ($\gg 5$~standard deviations) below the limit required for the presence of the entanglement ($-0.322 \pm 0.009$ for the {\sc POWHEG}+{\sc PYTHIA}~8 prediction). This means that ATLAS observes the entanglement.
It should be noted that the {\sc POWHEG}+{\sc PYTHIA} and {\sc POWHEG}+{\sc HERWIG} generators give different predictions. This was investigated and the size of the observed difference is consistent with changing the method of shower ordering. The deeper understanding would require further investigations. Also, the data do not agree with the prediction in the low $m(\ttbar)$ region. It is important to note that close to the threshold, non-relativistic QCD processes, such as Coulomb bound state effects, affect the production of \ttbar\ events  and are not accounted for in the simulation.

\section{Conclusion}
The ATLAS experiment studied the \ttbar\ pair spin correlations extensively. The full spin-correlation matrix was measured in the LHC Run 1 data taking.  
Recently, the spin correlations were shown to provide the information on the quantum entanglement and the ATLAS experiment observed the quantum entanglement in \ttbar\ pair production using the full Run 2 dataset.  
The \ttbar\ modeling is a limiting factor in a couple of areas for the ATLAS entanglement measurement. Therefore it is hoped this measurement will stimulate a progress in this area.
There are up to about 20 times more data expected with the full LHC program. Hopefully, this is therefore just a beginning of an era of the quantum information measurements at the LHC.

\section*{Acknowledgements}


\paragraph{Funding information}
This work was supported by EU and MEYS Project: \\ \mbox{FORTE: CZ.02.01.01/00/22\_008/0004632}.\\
The work was possible thanks to the support from the project: CERN-CZ LM2023040.

\bibliography{SciPost_Proceedings_TOP2024_Lysak.bib}

\begin{thebibliography}{10}
\providecommand{\url}[1]{\texttt{#1}}
\providecommand{\urlprefix}{URL }
\expandafter\ifx\csname urlstyle\endcsname\relax
  \providecommand{\doi}[1]{doi:\discretionary{}{}{}#1}\else
  \providecommand{\doi}{doi:\discretionary{}{}{}\begingroup
  \urlstyle{rm}\Url}\fi
\providecommand{\eprint}[2][]{\url{#2}}

\bibitem{ParticleDataGroup:2024cfk}
S.~Navas \emph{et~al.},
\newblock \emph{{Review of particle physics}},
\newblock Phys. Rev. D \textbf{110}(3), 030001 (2024),
\newblock \doi{10.1103/PhysRevD.110.030001}.

\bibitem{Bigi:1986jk}
I.~I.~Y. Bigi, Y.~L. Dokshitzer, V.~A. Khoze, J.~H. Kuhn and P.~M. Zerwas,
\newblock \emph{{Production and decay properties of ultra-heavy quarks}},
\newblock Phys. Lett. B \textbf{181}, 157 (1986),
\newblock \doi{10.1016/0370-2693(86)91275-X}.

\bibitem{Afik:2020onf}
Y.~Afik and J.~R. M.~n. de~Nova,
\newblock \emph{{Entanglement and quantum tomography with top quarks at the
  LHC}},
\newblock Eur. Phys. J. Plus \textbf{136}(9), 907 (2021),
\newblock \doi{10.1140/epjp/s13360-021-01902-1},
\newblock \eprint{2003.02280}.

\bibitem{Afik:2022kwm}
Y.~Afik and J.~R. M.~n. de~Nova,
\newblock \emph{{Quantum information with top quarks in QCD}},
\newblock Quantum \textbf{6}, 820 (2022),
\newblock \doi{10.22331/q-2022-09-29-820},
\newblock \eprint{2203.05582}.

\bibitem{ATLAS:2008xda}
{ATLAS Collaboration},
\newblock \emph{{The ATLAS Experiment at the CERN Large Hadron Collider}},
\newblock JINST \textbf{3}, S08003 (2008),
\newblock \doi{10.1088/1748-0221/3/08/S08003}.

\bibitem{TOPQ-2011-11}
{ATLAS Collaboration},
\newblock \emph{{Observation of spin correlation in \(t \bar{t}\) events from
  \(pp\) collisions at \(\sqrt{s} = 7\,\text{TeV}\) using the ATLAS detector}},
\newblock Phys. Rev. Lett. \textbf{108}, 212001 (2012),
\newblock \doi{10.1103/PhysRevLett.108.212001},
\newblock \eprint{1203.4081}.

\bibitem{TOPQ-2013-01}
{ATLAS Collaboration},
\newblock \emph{{Measurements of spin correlation in top--antitop quark events
  from proton--proton collisions at \(\sqrt{s} = 7\,\text{TeV}\) using the
  ATLAS detector}},
\newblock Phys. Rev. D \textbf{90}, 112016 (2014),
\newblock \doi{10.1103/PhysRevD.90.112016},
\newblock \eprint{1407.4314}.

\bibitem{TOPQ-2013-10}
{ATLAS Collaboration},
\newblock \emph{{Measurement of the correlation between the polar angles of
  leptons from top quark decays in the helicity basis at \(\sqrt{s} =
  7\,\text{TeV}\) using the ATLAS detector}},
\newblock Phys. Rev. D \textbf{93}, 012002 (2016),
\newblock \doi{10.1103/PhysRevD.93.012002},
\newblock \eprint{1510.07478}.

\bibitem{TOPQ-2014-07}
{ATLAS Collaboration},
\newblock \emph{{Measurement of Spin Correlation in Top--Antitop Quark Events
  and Search for Top Squark Pair Production in \(pp\) Collisions at \(\sqrt{s}
  = 8\,\text{TeV}\) Using the ATLAS Detector}},
\newblock Phys. Rev. Lett. \textbf{114}, 142001 (2015),
\newblock \doi{10.1103/PhysRevLett.114.142001},
\newblock \eprint{1412.4742}.

\bibitem{TOPQ-2015-13}
{ATLAS Collaboration},
\newblock \emph{{Measurements of top quark spin observables in \(t\bar{t}\)
  events using dilepton final states in \(\sqrt{s} = 8\,\text{TeV}\) \(pp\)
  collisions with the ATLAS detector}},
\newblock JHEP \textbf{03}, 113 (2017),
\newblock \doi{10.1007/JHEP03(2017)113},
\newblock \eprint{1612.07004}.

\bibitem{TOPQ-2016-10}
{ATLAS Collaboration},
\newblock \emph{{Measurements of top-quark pair spin correlations in the
  \(e\mu\) channel at \(\sqrt{s} = 13\,\text{TeV}\) using \(pp\) collisions in
  the ATLAS detector}},
\newblock Eur. Phys. J. C \textbf{80}, 754 (2020),
\newblock \doi{10.1140/epjc/s10052-020-8181-6},
\newblock \eprint{1903.07570}.

\bibitem{ATLAS:2023fsd}
{ATLAS Collaboration},
\newblock \emph{{Observation of quantum entanglement with top quarks at the
  ATLAS detector}},
\newblock Nature \textbf{633}(8030), 542 (2024),
\newblock \doi{10.1038/s41586-024-07824-z},
\newblock \eprint{2311.07288}.

\bibitem{Bernreuther:2015yna}
W.~Bernreuther, D.~Heisler and Z.-G. Si,
\newblock \emph{{A set of top quark spin correlation and polarization
  observables for the LHC: Standard Model predictions and new physics
  contributions}},
\newblock JHEP \textbf{12}, 026 (2015),
\newblock \doi{10.1007/JHEP12(2015)026},
\newblock \eprint{1508.05271}.

\bibitem{Behring:2019iiv}
A.~Behring, M.~Czakon, A.~Mitov, A.~S. Papanastasiou and R.~Poncelet,
\newblock \emph{{Higher order corrections to spin correlations in top quark
  pair production at the LHC}},
\newblock Phys. Rev. Lett. \textbf{123}(8), 082001 (2019),
\newblock \doi{10.1103/PhysRevLett.123.082001},
\newblock \eprint{1901.05407}.

\end{thebibliography}


\end{document}